\documentclass[twocolumn,pre,superscriptaddress]{revtex4}
\usepackage{graphicx}

\newcommand{\omegav}{\mbox{\boldmath$\omega$}}
\newcommand{\bmi}[2]{\begin{minipage}[#1]{#2 \linewidth}}
\newcommand{\emi}{\end{minipage}}

\begin{document}

\title{Depletion forces between non-spherical objects}
\author{P.-M. K\"onig}
\affiliation{Max-Planck-Institut f{\"u}r Metallforschung,
Heisenbergstr. 3, D-70569 Stuttgart, Germany\\}
\affiliation{Institut f{\"u}r Theoretische und Angewandte Physik,
Universit{\"a}t Stuttgart, Pfaffenwaldring 57, D-70569 Stuttgart, Germany\\}
\author{R. Roth}
\affiliation{Max-Planck-Institut f{\"u}r Metallforschung,
Heisenbergstr. 3, D-70569 Stuttgart, Germany\\}
\affiliation{Institut f{\"u}r Theoretische und Angewandte Physik,
Universit{\"a}t Stuttgart, Pfaffenwaldring 57, D-70569 Stuttgart, Germany\\}
\author{S. Dietrich}
\affiliation{Max-Planck-Institut f{\"u}r Metallforschung,
Heisenbergstr. 3, D-70569 Stuttgart, Germany\\}
\affiliation{Institut f{\"u}r Theoretische und Angewandte Physik,
Universit{\"a}t Stuttgart, Pfaffenwaldring 57, D-70569 Stuttgart, Germany\\}

\begin{abstract}
We extend the insertion approach for calculating depletion potentials to the
case of non-spherical solutes. Instead of a brute-force calculation we suggest
to employ the recently developed curvature expansion of density profiles close
to complexly shaped walls. The approximations introduced in the calculation by
the use of the curvature expansion and of weight functions for non-spherical
objects can be tested independently. As an application for our approach we
calculate and discuss the depletion potential between two hard oblate
ellipsoids in a solvent of hard spheres. For this system we calculate the
entropic force and torque acting on the objects.
\end{abstract}

\maketitle

\section{Introduction}

If macromolecules such as colloids are immersed in a solvent of smaller
particles, the difference in size makes it useful to describe this mixture in
terms of effective interactions by integrating out the degrees of freedom of
the solvent. The resulting interactions between particles of the remaining
larger component are often referred to as depletion interactions. Such
depletion forces have been studied in detail both theoretically
\cite{Asakura54,Asakura58,Biben96,Dickman97,Roth99,Roth00} and 
experimentally \cite{Rudhardt98,Bechinger99,Helden03}. Most theoretical
approaches are based on a brute-force approach in which the solutes are
frozen in a given configuration and thereby turned into an external field for
the solvent \cite{Biben96,Dickman97,Kinoshita02,Kinoshita04,Kinoshita06}. From 
the inhomogeneous structure of the solvent in the external field due to two
fixed solute particles one can calculate the solvent mediated effective force
acting on the solutes. This brute-force approach turns out to be very time
consuming because for each separation and orientation, for which one wants to
determine the depletion force, the inhomogeneous solvent distribution has to
be calculated anew.  

The insertion approach to depletion potentials \cite{Roth00} differs in
character in that there only one solute particle is fixed. The advantage is
that the calculation of the solvent density profile in the external field of a
single solute is usually much simpler and hence computational less
demanding. The potential of the depletion force can be determined from the
solvent density profile close to one solute by inserting the second solute
into the system using the potential distribution theorem \cite{Henderson83}. 
However, for the insertion step a theoretical description of a mixture
consisting of solute and solvent particles is required. 

So far, most studies of depletion forces have been focused on rather simple
geometries, such as the force between a big sphere and a planar wall or
between two big spheres in a solvent of small spheres. For these symmetric 
systems the depletion potential depends on the sphere-wall or the 
sphere-sphere separation as the only parameter characterizing the 
configuration. In addition, for these geometries the use of the aforementioned 
insertion approach is facilitated by the availability of reliable density 
functional theories for hard-sphere mixtures. 

In colloidal mixtures one generally encounters more complex particle shapes
both of the solute and solvent particles. However, the understanding of
depletion potentials for non-spherical objects is still rudimentary. The
corresponding calculations are much more challenging because in these cases
the depletion potential depends not only on the separation between the solutes
but also on their relative orientation. 

A first extension beyond the mixture of spheres is the case of
spherical solutes immersed in a solvent of non-spherical particles. Depletion
agents such as thin rods \cite{Mao95,Mao97,Yaman98,Roth03} or infinitely thin
platelets \cite{Oversteegen03,Harnau04,Harnau06} can generate big depletion
effects even at rather low solvent concentrations \cite{Helden03}. The
depletion potential in these cases can be calculated in the limit of small
solvent densities so that correlations among the depletion agents are
small. Similar in spirit is also the calculation of the depletion force
between spheres in a solvent of a liquid-crystal in its nematic phase
\cite{Schoot00,Schoot00b}, for which the strong correlations between 
particles of the liquid crystal are taken into account effectively by
reducing their orientational degrees of freedom. The strength of the depletion
interaction is then estimated by excluded volume calculations, following the
ideas of Asakura and Oosawa \cite{Asakura54,Asakura58}.

A second, more complicated situation is the one studied in
Ref.~\cite{Roth02a}, where one spherocylinder immersed in a spherical solvent
close to a planar wall was considered. These calculations employed the
insertion approach. The resulting depletion potential depends not only on the
separation of the solute from the wall but also on its orientation. Hence in
addition to the depletion force an entropic torque acts on the solute. 

There are recent studies of depletion forces between two spherocylinders in a
solvent of spheres which overcome the Asakura-Oosawa approximation
\cite{Asakura54,Asakura58}. Using a three-dimensional
integral-equation theory, Kinoshita \cite{Kinoshita04,Kinoshita06} 
showed that the corresponding depletion potential displays a rich behavior and
depends sensitively on the path along which the spherocylinders approach each
other. Similar findings were reported in a simulation study, in which the
depletion potential was determined by the acceptance ratio method 
\cite{Li05}.

Here we extend the DFT insertion approach to depletion potentials
\cite{Roth00} to the case of non-spherical objects. In Sec.~\ref{sec:theory} 
we recall the basic theory and highlight how the geometry of the solutes can
be accounted for. We test the new elements of the theory in
Sec.~\ref{sec:test}. As an application we discuss the depletion potential
between two ellipsoids in Sec.~\ref{sec:application}. We conclude in
Sec.~\ref{sec:conclusion}. 

\section{Theory}
\label{sec:theory}

We follow the versatile and successful approach to calculating depletion
potentials between two objects, $a$ and $b$, immersed in a solvent 
within the framework of density function theory (DFT) \cite{Roth00},
which is referred to as the insertion approach. To this end we fix one of the
objects, say $a$, at the origin at a given orientation so that it acts as an
external potential for the solvent particles. In response to this external
potential, the {\em s}olvent particles aquire an inhomogeneous equilibrium
number density distribution $\rho_s({\bf r})$. In the fluid phase, the case
we are interested in here, $\rho_s({\bf r})$ shares the spatial symmetry with
that of object $a$. If object $a$ is a sphere, the density distribution
$\rho_s({\bf r})$ possesses also spherical symmetry. In a second step we insert
the second object, denoted by $b$, into the inhomogeneous solvent at position
${\bf r}$ and with relative orientation ${\omegav}$. As a result of this
insertion the grand potential $\Omega({\bf r},{\omegav})$ of the system
changes. The depletion potential is given by \cite{Roth00}
\begin{equation}
W({\bf r},{\omegav}) = \Omega({\bf r},{\omegav}) - 
\Omega({\bf r}\to\infty,{\omegav}),
\end{equation}
which can be re-written in terms of the one-body direct correlation
function $c_b^{(1)}({\bf r},{\omegav})=-\beta\delta {\cal F}_{ex}/
\delta \rho_b({\bf r},{\omegav})$ \cite{Roth00}:
\begin{equation} 
\label{dep}
\beta W({\bf r},{\omegav}) = c_b^{(1)}({\bf r}\to\infty,{\omegav})-
c_b^{(1)}({\bf r},{\omegav}).
\end{equation}
For the numerical calculation of $c_b^{(1)}({\bf r},{\omegav})$ two 
challenges have to be overcome: (i) the {\em accurate} calculation of the
density profile $\rho_s({\bf r})$ of solvent particles around an object of
complex shape, and (ii) the insertion of a non-spherical object into an
inhomogeneous solvent of spheres. Since both of these problems require
non-standard approaches, in the following we shall pay special attention to
them. 

Although our approach is flexible and can treat also soft solvent-solvent and
solute-solvent interactions, in the following we shall restrict our
considerations to the case of hard-core interactions. The solvent is
represented by a hard-sphere fluid characterized by its radius $R$ and bulk
density $\rho_s$ or bulk packing fraction $\eta_s=4 \pi R^3 \rho_s/3$. The
solute-solvent interaction is infinitely repulsive in the case of overlap and
zero otherwise. 

\subsection{Density profiles and weighted densities}

\subsubsection{Curvature expansion}

Concerning the first issue of calculating the density profile around
non-spherical objects we apply the recently suggested and successfully tested
curvature expansion of the density profile \cite{Koenig05}. To this end
we introduce normal coordinates. Any point ${\bf r}$ outside the fixed object 
$a$ can be reached from ${\bf R}$ as the point closest to ${\bf r}$ on the
parallel surface of the object, where the density profile $\rho_s({\bf r})$
vanishes discontinuously. This particular surface is special to the case of
hard-core solute-solvent interaction; however, it is possible to employ any
parallel surface as long as all the calculations are in line with this
definition of ${\bf R}$.

For the vector connecting the points ${\bf R}$ and ${\bf r}$ one has
${\bf R}-{\bf r}= u {\bf n(R)}$, where ${\bf n(R)}$ is the 
unit vector normal to the parallel surface at point ${\bf R}$ and $u$ is the
normal distance. At point ${\bf R}$ the parallel surface exhibits two
principle radii of curvature $R_1$ and $R_2$ leading to the dimensionless mean
and Gaussian curvatures $H({\bf R}) = (R/R_1+R/R_2)/2$ and $K({\bf R}) =
R^2/(R_1 R_2)$, respectively.

As an ansatz for the density profile we employ \cite{Koenig05}
\begin{eqnarray} \label{cerho}
\rho_s({\bf r}) = \rho_s^P(u) + H({\bf R}) \rho_s^H(u) + K({\bf R}) \rho_s^K(u)
\nonumber \\ + H({\bf R})^2 \rho_s^{H^2}(u) + H({\bf R})^3 \rho_s^{H^3}(u)
\nonumber \\ +H({\bf R}) K({\bf R})\rho_s^{H K}(u)\dots,
\end{eqnarray}
which factorizes the local geometry of object $a$ at position ${\bf R}$,
specified by the local mean and Gaussian curvature $H({\bf R})$ and 
$K({\bf R})$, respectively, and the structure of the solvent via the
coefficient functions $\rho_s^\xi(u)$ with $\xi=P,H,K,\dots$. This separation 
of geometry and the coefficient functions allows one to infer $\rho_s^\xi(u)$
from simple geometries with high symmetry, such as a fluid close to planar,
spherical, or cylindrical walls. The functions $\rho_s^\xi(u)$ have been
determined in Ref.~\cite{Koenig05}. There the direct comparison between a
density profile predicted by the curvature expansion, based on these
coefficient functions and obtained within DFT, with results from Monte-Carlo
simulations has demonstrated the high accuracy of this approach. Note that the
curvature expansion of the density profile, Eq.~(\ref{cerho}), implicitly
assumes that the curvature of the fixed object varies smoothly on its the
surface. A sharp edge would be problematic because the mean curvature jumps
from zero to a non-vanishing value at the edge. 

\subsubsection{Contact density on spherocylinders}

On the surface of a spherocylinder both the mean and the Gaussian curvature
vary discontinuously where the spherical cap meets the cylinder. Also this
discontinuity of the curvatures cannot be captured fully by the curvature
expansion.

\begin{figure}
\includegraphics[width=0.9\linewidth]{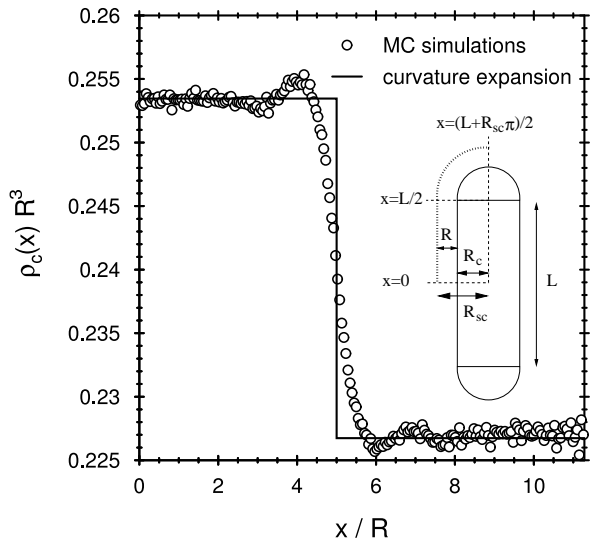}
\caption{\label{fig:compare}
The contact density $\rho_c(x)$ of a fluid of hard spheres with radius $R$ and
packing fraction $\eta_s=0.3$ at a hard spherocylinder with length $L=10 R$ and
radius $R_c=3 R$ so that the contact density occurs at $R_{sc}=4 R$. The path
on the parallel surface of contact of the spherocylinder is parameterized by
$x$, as is indicated by the dotted line in the inset. Note that the inset is 
not drawn to scale. According to the curvature expansion the contact value 
$\rho_c(x)$ jumps at $x=L/2$, where the cylindrical and the spherical parts 
meet, as shown by the full line. The symbols denote data from Monte Carlo 
simulations for $\eta_s=0.3$.}
\end{figure}

In order to analyze the reliability of the curvature expansion we have
performed a Monte Carlo simulation of a hard-sphere fluid exposed to a
spherocylinder. As parameters we have chosen $\eta_s=0.3$ for the packing
fraction of the fluid, $L=10 R$ for the length of the cylinder, and 
$R_{sc}=4 R$ for the radius of the parallel surface at which the density
profile discontinuously drops to zero. In this test we focus on the contact
density $\rho_c(x)$ as the most sensitive quantity, where $x$ parameterizes a
path along the surface as depicted in the inset of Fig.~\ref{fig:compare}. The
density profile away from contact decays towards the bulk density and we 
verified that that the effects of the discontinuity of the curvature decreases
with increasing normal distance from the spherocylinder.

The dimensionless curvatures on the cylindrical part of the surface are
$H=R/(2 R_{sc})$ and $K=0$, while they are $H=R/R_{sc}$ and  $K=(R/R_{sc})^2$
on the spherical caps. Accordingly, the curvature expansion predicts a jump in
the contact value of the density profile where the spherical caps meet the
cylinder. The result of the curvature expansion is plotted as the full line in
Fig.~\ref{fig:compare}. Note that in Eq.~(\ref{cerho}) only the three terms 
$\rho_s^P(u)$, $\rho_s^H(u)$, and $\rho_s^K(u)$ have a non-vanishing
contact value and hence contribute to $\rho_c(x)$ \cite{Koenig04,Koenig05}.

In contrast to the jump of the contact density between constant values as
predicted by the curvature expansion, in the computer simulations we find a
smooth, slightly oscillatory transition between the contact densities at the
cylindrical and the spherical part of the spherocylinder. In
Fig.~\ref{fig:compare} the simulation data for $\rho_c(x)$ are shown as 
symbols. Interestingly, the spatial region of deviation between the simulation
data and the prediction of the curvature expansion is relatively narrow.

\subsubsection{Curvature expansion of the free energy density and of weighted
  densities}

Although the density profile $\rho_s({\bf r})$, as given by Eq.~(\ref{cerho}),
is in principle sufficient for the calculations we intend to perform, it is
numerically more efficient to exploit the particular form of
fundamental-measure theory (FMT) functionals \cite{Rosenfeld89}. Within FMT
the one-body direct  correlation function is given by
\begin{equation}
\label{cb1}
c_b^{(1)} = - \sum_\alpha \frac{\partial \Phi}{\partial n_\alpha}
\otimes w_\alpha^b,
\end{equation}
where $\Phi$ is the excess free energy density and $\Psi_\alpha\equiv\partial
\Phi/\partial n_\alpha$ depends on $\rho_s({\bf r})$ in a complicated, 
nonlinear way. The convolution product is denoted as $\otimes$. Instead of
calculating $\Psi_\alpha$ from $\rho_s({\bf r})$ directly, we argue that
$\Psi_\alpha({\bf r})$ can be equivalently expanded in terms of powers of $H$
and $K$ and therefore can be written as
\begin{equation} \label{cepsi}
\Psi_\alpha({\bf r}) = \Psi_\alpha^P(u) + H({\bf R}) \Psi_\alpha^H(u) +
K({\bf R}) \Psi_\alpha^K(u) + \dots .
\end{equation}
One can adopt the point of view that curvature expansions such as those given
in Eqs.~(\ref{cerho}) and (\ref{cepsi}) are merely approximations of the
functions $\rho_s({\bf r})$ or $\Psi({\bf r})$ that take the shape of the
external potential into account in an efficient way. Within this line of
arguments there is nothing special about the density distribution or any
other function entering the DFT. Even auxiliary functions such as the weighted
densities $n_\alpha({\bf r})$ and $\Psi_\alpha({\bf r})$ have a curvature
expansion ensuring that the output of the DFT calculation has the form of
Eq.~(\ref{cerho}).

A more systematic, albeit more involved, point of view, which we present
here only as a sketch, starts with an approach similar in spirit to the one
that leads to Eq.~(\ref{cerho}) \cite{Koenig05} as a suitable form for the
density profile. Here, however, we analyze the weighted densities, defined as
\begin{equation}
n_\alpha({\bf r}) = \int d {\bf r}'\rho_s({\bf r}') w_\alpha({\bf r},{\bf r}'),
\end{equation}
close to planar, spherical, and cylindrical walls. Our results suggest that
analogous to the density profile [Eq.~(\ref{cerho})] also the weighted
densities can be expanded in terms of powers of the curvatures $H$ and $K$. As
in Ref.~\cite{Koenig05} we can determine uniquely the coefficient functions
$n_\alpha^\xi$ for $\xi=P,H,K,H^2,H K,H^3$ up to third order in the inverse of
the radii of curvatures inferred from planar, spherical, and cylindrical
geometries. In order to obtain higher order contributions one would have to
consider more complex wall shapes. The numerical accuracy of weighted
densities calculated at more complex walls is, however, unsatisfactory and
practically prevents a reliable decomposition into coefficient functions. 
Since we are interested in calculating the depletion potential between two 
{\em big} non-spherical objects immersed in a solvent of {\em small} spheres, 
the coefficient functions which we have determined are sufficient. For our
approach to be quantitatively reliable, curvatures of the surface of the
non-spherical objects should always be sufficiently small.  

We therefore use as an ansatz for the weighted densities, following 
Ref.~\cite{Koenig05},
\begin{equation} \label{cena}
n_\alpha({\bf r}) = n_\alpha^P(u) + H({\bf R}) n_\alpha^H(u) + 
K({\bf R}) n_\alpha^K(u) + \dots,
\end{equation}
which we can insert into the free-energy density $\Phi$ or its derivative
with respect to $n_\alpha$, i.e., $\Psi_\alpha$. Both $\Phi$ and $\Psi_\alpha$
are highly nonlinear functions of $n_\alpha$. However, by Taylor expanding
$\Phi$ or $\Psi_\alpha$ into powers of $n_\alpha$ and by re-arranging terms one
can see immediately that the curvature expansions of $\Psi_\alpha$ follow
directly from Eq.~(\ref{cena}).

\subsection{Insertion of non-spherical objects}

We now turn to the second part of the present problem, i.e., the calculation
of the insertion free energy [Eq.~(\ref{dep})] of the non-spherical object
$b$. Note that the change in the grand potential due to the insertion of
object $b$ into a {\em homogeneous} bulk fluid at ${\bf r}\to\infty$ can be
described by using the morphometric approach \cite{Koenig04,Koenig05}. In the
bulk, the change in grand potential of the system cannot depend on the
orientation of the inserted object, which simplifies the problem
somewhat. Furthermore, it was shown that the problems even simplify further
due to the separation of the geometry and the shape independent thermodynamical
coefficients. In the morphometric approach, the insertion free energy of
object $b$ in a bulk fluid can be written as \cite{Koenig04,Koenig05}
\begin{equation}
-\beta^{-1} c_b^{(1)}({\bf r}\to\infty,{\omegav}) = p V_b + \sigma A_b +
 \kappa C_b +  \bar{\kappa} X_b,
\end{equation}
where $p$, $\sigma$, $\kappa$, and $\bar{\kappa}$ are the pressure, the planar
wall surface tension, and two bending rigidities, respectively, which depend on
the state of the bulk fluid and the interaction between the fluid and object
$b$. These coefficients can be obtained in simple geometries. The corresponding
geometrical measures  $V_b$, $A_b$, $C_b$, and $X_b$ describing the shape of
object $b$ are the volume, the surface area, and the integrated (over the
surface area) mean and Gaussian curvature, respectively. 

In order to calculate the convolutions in Eq.~(\ref{cb1}) the weight functions
$w_\alpha^b$ for a non-spherical object are required. For this problem we
employ Rosenfeld's formulation of fundamental measure theory generalized to
convex hard bodies \cite{Rosenfeld94,Rosenfeld95}. There are four scalar
weight functions:
\begin{equation}
w_3^b({\bf r}) = \Theta(|{\bf r}-{\bf R}_b(\theta,\phi)|) ,
\end{equation}
which defines the volume $V_b$ of object $b$,
\begin{equation}
w_2^b({\bf r}) = \delta({\bf r}-{\bf R}_b(\theta,\phi)),
\end{equation}
which defines the surface area $A_b$ of $b$,
\begin{equation}
w_1^b({\bf r}) = \frac{H({\bf r}) w_2^b}{4 \pi},
\end{equation}
which defines the integrated (over the surface) mean curvature $C_b$ of $b$, 
and
\begin{equation}
w_0^b({\bf r}) = \frac{K({\bf r}) w_2^b}{4 \pi},
\end{equation}
which defines the integrated (over the surface) Gaussian curvature or Euler 
characteristics $X_b$ of object $b$. Besides the scalar weight functions, 
which represent the geometrical properties of $b$, there are two additional 
vector-like weight functions which are required for the de-convolution of the 
Mayer-$f$ function describing the interaction between non-spherical particles. 
The vector-like weight functions are  given by  
\begin{equation}
{\bf w}_2({\bf r}) = -\nabla w_3^b({\bf r}) = {\bf n}_b({\bf r})
\delta({\bf r}-{\bf R}_b(\theta,\phi)),
\end{equation}
where ${\bf n}_b({\bf r})$ is the unit vector of the surface normal at point
${\bf r}$, and 
\begin{equation}
{\bf w}_1({\bf r}) = \frac{H({\bf r}){\bf w}_2({\bf r})}{4 \pi}  .
\end{equation}
In a bulk system the vector-like weighted densities 
${\bf n}_2({\bf r})$ and ${\bf n}_1({\bf r})$ vanish. These weight functions
have been employed successfully in the calculation of the depletion potential
between a hard spherocylinder and a planar hard wall \cite{Roth02a}. 

With Eqs.~(\ref{dep}) and (\ref{cb1}) we can now calculate the depletion
potential $W({\bf r},{\omegav})$.

\subsection{Force and torque}

From the knowledge of the depletion potential $W({\bf r},{\omegav})$ it is
possible to determine the entropic force and the entropic torque
\cite{Roth02a} acting on object $b$ with orientation ${\omegav}$ at a given
position ${\bf r}$. If object $b$ is translated by an infinitesimal
vector $\delta {\bf r}$, while keeping its orientation fixed, the depletion
potential changes by $\delta W = - {\bf F} \cdot \delta {\bf r}$, which
defines the depletion force
\begin{equation}
{\bf F}({\bf r},{\omegav}) = - \frac{\partial}{\partial {\bf r}} 
W({\bf r},{\omegav}).
\end{equation}
The torque acting on the object $b$ can be calculated by a similar
consideration, rotating object $b$ by an infinitesimal angle $\delta
\omegav$. The direction of $\delta \omegav$ is parallel to the axis of
rotation and its modulus specifies the angle of rotation. If one keeps the
center of $b$ fixed at ${\bf r}$ and performs a rotation by $\delta \omegav$,
the depletion potential changes by  $\delta W$, so that
$\delta W = - {\bf M} \cdot \delta \omegav$. Therefore we can write the
entropic torque as 
\cite{Roth02a}
\begin{equation}
\label{torque}
{\bf M}({\bf r},{\omegav}) = - \frac{\partial}{\partial {\omegav}} 
W({\bf r},{\omegav}).
\end{equation}

\section{Test: depletion potentials between ellipsoids and spheres}
\label{sec:test}

\begin{figure}
\includegraphics[width=0.9\linewidth]{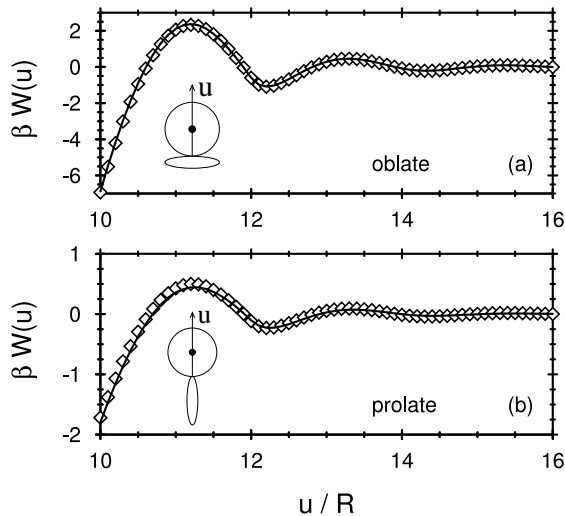}
\caption{\label{fig:test} 
Depletion interaction between an oblate (a) [prolate (b)] ellipsoid
with half-axes  $(10,10,4)R$ [$(4,4,10)R$] and a sphere with $R_a=10R$. The
center of the inserted sphere approaches the north pole of the fixed ellipsoid
normal to the surface. Both particles  touch if $u=R_a$. The symbols denote
results obtained by approach (1), for which we employ FMT for convex objects,
and the lines denote corresponding results from approach (2), for which we
use the curvature expansion. As both approaches involve approximations of very
different nature, the excellent agreement is very likely to be due to the
fact that the systematic error is very small in both approaches. We verified
this observation also for ellipsoids with different half-axes. For these data 
the depletion agent is a fluid of hard spheres with radius $R$ modeled via the 
White-Bear version of FMT. The bulk packing fraction is $\eta_s=0.3$. Note 
that the insets here and in the following figures are not drawn to scale.} 
\end{figure}

Before we apply the above formalism to the calculation of the depletion
potential between two non-spherical objects, we perform a test that enables us
to estimate the errors introduced into the numerical calculation through the
approximations we have made. One source of error is the use of the curvature
expansion of the density profile $\rho_s({\bf r})$ [Eq.~(\ref{cerho})] and of
the derivatives $\Psi_\alpha({\bf r})$ of the excess free-energy density
[Eq.~(\ref{cepsi})]. It introduces an approximation because we have to truncate
the expansion after the third order in the inverse radii of curvature. Another
source of error is the application of the FMT weight functions for
non-spherical objects, which introduces a different approximation. The fact
that these two approximations are very distinct in nature allows us to perform
a stringent numerical test.

To this end we calculate the depletion potential between one big sphere and
one big ellipsoid in a solvent of small spheres. In an {\em exact} treatment
the depletion potential between these two objects would depend only on their
relative position and orientation. In the numerical implementation of the
insertion method we  can follow two different routes which make independent
use of the different approximations. We can choose to fix either the sphere and
insert the ellipsoid or fix the ellipsoid and insert the sphere. By the choice
of particle fixed at the origin we decide about the symmetry of the external
potential and how to calculate the density distribution of the solvent. 

If we choose to fix the big sphere the calculation of the density profile
$\rho_s({\bf r})=\rho_s(r)$ is straightforward and, using the spherical
symmetry of the problem, it was established that the results agree extremely
well with, e.g., Monte Carlo simulations. Along this route, the main
approximation for the calculation of $W({\bf r},{\omegav})$ stems from using
the weight functions for the ellipsoid. 

Note that the resulting depletion potential $W(r,{\omegav})$ depends on
both the orientation $\omegav$ of the ellipsoid relative to the vector {\bf r}
connecting the centers of the ellipsoid and the sphere and the distance $r$ 
between the ellipsoid and the sphere. For this test, however, we fix the
orientation and consider the approach between the ellipsoid and the sphere
along the surface normal of the sphere for the chosen orientation of the
ellipsoid. In Fig.~\ref{fig:test} we show the depletion potential between a
big sphere, denoted as object $a$, with radius $R_a=10 R$ and an oblate
ellipsoid with half-axes $(10,10,4) R$ (a), and between a sphere with radius
$R_a=10 R$ and a prolate ellipsoid with half axes ($4,4,10) R$ (b). In both
cases the solvent is a fluid of small spheres with a packing fraction
$\eta_s=0.3$, which we model by the White Bear version of FMT
\cite{Roth02b,Wu02}. The insets in Fig.~\ref{fig:test} depict the orientation
between the ellipsoids and the sphere chosen in the calculation. The symbols
denote the results obtained via the first route, corresponding to a fixed 
sphere. 

Along the second route, we fix the ellipsoid at the origin so that it acts as 
an external potential for the solvent spheres. Now we employ the curvature
expansion [Eq.~(\ref{cepsi})] in order to evaluate the derivatives of the
excess free energy density $\Psi_\alpha({\bf r})$. The weight functions we
need in order to describe the insertion of the big sphere in Eq.~(\ref{cb1})
are well tested and are known to be accurate \cite{Roth00}. The results for
the depletion potentials along the same paths between the sphere and the
oblate and prolate ellipsoid obtained from this route are shown in
Fig.~\ref{fig:test} as lines.

We find that the results obtained from both routes agree extremely well, which 
provides confidence in the reliability of the approximations and the numerical
approach. Only in the case of the prolate ellipsoid, for which the curvatures
are considerably higher than in the case of the oblate ellipsoid, we find some
deviations between the two routes close to the first potential
barrier. However, these deviations are very small.

Close to contact between the sphere and the ellipsoids, one can employ
arguments based on considerations about the overlap of excluded volumes
\cite{Asakura54,Asakura58,Roth99}. In the case of high curvature one expects
that the contact value of the depletion potential is considerably reduced
compared to cases of low curvature. This expectation is confirmed by our
results. In addition to the contact value, we find that the amplitude of the
oscillations of the depletion potential for the oblate ellipsoid
[Fig.~\ref{fig:test}(a)] is larger than the one for the prolate ellipsoid
[Fig.~\ref{fig:test}(b)].

We conclude from the results of this test that both the curvature expansion of
the functions $\Psi_\alpha({\bf r})$ and the insertion of a non-spherical
particle into an inhomogeneous solvent of small spheres work reliably.  

\section{Application: depletion potentials between two ellipsoids}
\label{sec:application}

We can now turn to the calculation of the depletion potential between {\em two}
non-spherical, convex objects. In order to accomplish this we have to combine
both steps mentioned and tested above. First we fix one of the two
non-spherical objects and thus turn it into an external potential for the
solvent of small spheres. The structure of the resulting inhomogeneous solvent
density distribution at a given bulk density is captured by the curvature
expansion given by Eq.~(\ref{cepsi}). In the second step we employ the 
insertion of a non-spherical object.

\begin{figure}
\includegraphics[width=0.9\linewidth]{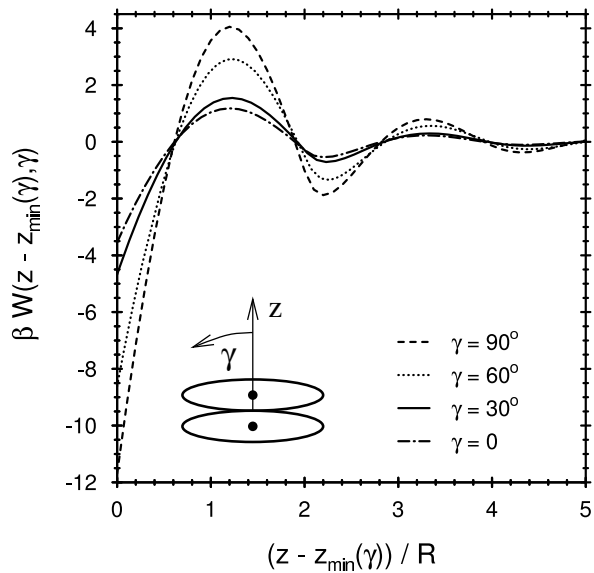}
\caption{\label{fig:twooblates} 
Depletion interaction between two oblate ellipsoids with half axes
$(10,10,4)R$. The inserted particle approaches the north pole of the first
ellipsoid perpendicular to its surface. $\gamma$ denotes the angle between the
z-axis and the large half-axis of the inserted ellipsoid (see the inset). If 
both ellipsoids are aligned (as drawn in the inset), i.e., $\gamma=90^\circ$, 
the minimal separation of the centers is $z_{min}=8 R$, while for $\gamma=0$ 
it is $z_{min}=14 R$; $z_{min}(\gamma=60^\circ)\approx 9.53 R$ and 
$z_{min}(\gamma=30^\circ)\approx 12.57 R$. The positions of the extrema and 
the zeros are basically independent of $\gamma$. The solvent is a fluid of
hard spheres with radius $R$ and bulk packing fraction $\eta_s=0.3$ and is
modeled via the White-Bear density functional.}  
\end{figure}

In order to illustrate our approach we calculate the depletion potential
between two hard oblate ellipsoids with half-axes $(10,10,4) R$ in a solvent
of hard spheres with radius $R$ and bulk packing fraction $\eta_s=0.3$. The
resulting depletion potential depends on both the relative orientation
$\omegav$ of the two ellipsoids and the vector ${\bf r}$ connecting their
centers and thus depends on 6 variables. Out of this high-dimensional parameter
space we select a few examples of paths along which we study the behavior of
the depletion potential. 

In the first example, the center of the inserted ellipsoid approaches the
north pole of the fixed particle along the surface normal, as shown in the
inset of Fig.~\ref{fig:twooblates}. For this path we vary also the relative
orientation $\omegav$ between the particles. Due to the symmetry of this
configuration the orientation between the oblate ellipsoids can be expressed
in terms of a single angle, which we denote by $\gamma$. If we choose
$\gamma=90^\circ$, the ellipsoids are parallel and the minimal separation of
their centers is $z_{min}=8 R$. At contact the overlap of excluded volume is
larger than for any other value of $\gamma$. Hence the contact value of the
depletion potential is most negative. For our choice of parameters we find
$\beta W(z_{min},\gamma=90^\circ)\simeq -12$ (see
Fig.~\ref{fig:twooblates}). In addition to the strong attraction close to
contact, the depletion potential displays a pronounced oscillatory structure
away from contact.

This oscillatory structure of the depletion potential reflects mainly the
properties of the solvent. The structure of the hard-sphere solvent normal to
the surface displays oscillatory, exponentially decaying packing effects. We
have shown previously \cite{Roth00} that beyond the first maximum the 
depletion potential between two spheres or between a sphere and a planar wall
decays in a closely related, exponentially damped oscillatory fashion. The
decay length and the wavelength of the oscillations in the depletion potential
are exactly the same as those in the decay of the bulk pair correlation
function. Only the amplitude of this decay and the phase of the oscillations
depend on the shape and the orientation of the two solutes. 

\begin{figure}
\includegraphics[width=0.9\linewidth]{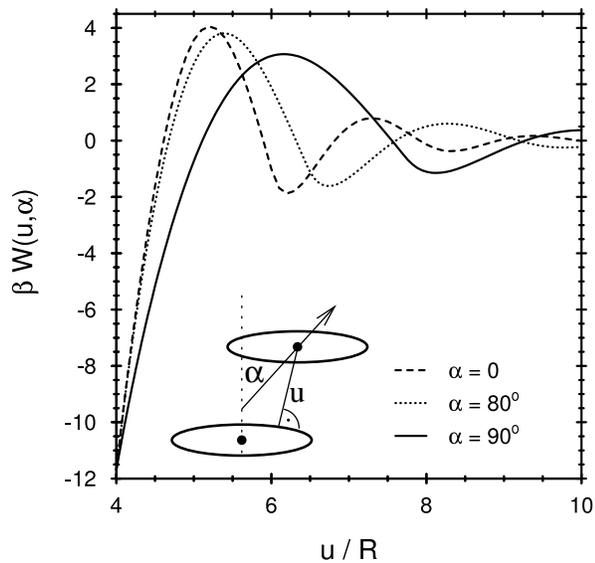}
\caption{\label{fig:oblatesangle}
Depletion potential for the same setup as in Fig.~\ref{fig:twooblates} with
fixed $\gamma=90^\circ$. The angle $\alpha$ characterizes the straight line
along which the center of the inserted ellipsoid is moved away from its contact
position at the north poles (see the inset). The abscissa measures the normal
(i.e., minimal) distance $u$ of the center of the inserted ellipsoid from the
surface of the fixed one.  For small $\alpha$ the depletion potentials almost
coincide due to the oblate shape $(10,10,4)R$ of the fixed ellipsoid.} 
\end{figure}

As we decrease the value of $\gamma$ from 90$^\circ$ to 0, thereby changing
the relative orientation from parallel to normal, we find that the contact
value as well as the potential barrier at $z-z_{min} \gtrsim R$ and the
amplitude of the oscillations decrease monotonically; the positions of
the extrema and of the zeros basically do not vary with $\gamma$.

In the second example the ellipsoids always approach each others north poles
in a parallel configuration. The corresponding path of the centers is a
straight line forming an angle $\alpha$ with the fixed direction of the small
axes (see the inset of Fig.~\ref{fig:oblatesangle}). In
Fig.~\ref{fig:oblatesangle}, by construction the path for $\alpha=0$ is
identical to the path with $\gamma=90^\circ$ in Fig.~\ref{fig:twooblates}. Upon
increasing $\alpha$, at first the resulting depletion potential changes only
slightly but for angles $\alpha\gtrsim 80^\circ$ it is possible to observe a
clear decrease in the wavelength of oscillation, which is most pronounced in
the case of $\alpha=90^\circ$. This, however, is not a contradiction to the
aforementioned universality of the oscillatory decay of the depletion
potential, because this path cuts through the three-dimensional oscillatory
structure of the solvent, which is organized {\em normal} to the surface of
the fixed solute, at the angle $\alpha$.

\begin{figure}
\includegraphics[width=0.9\linewidth]{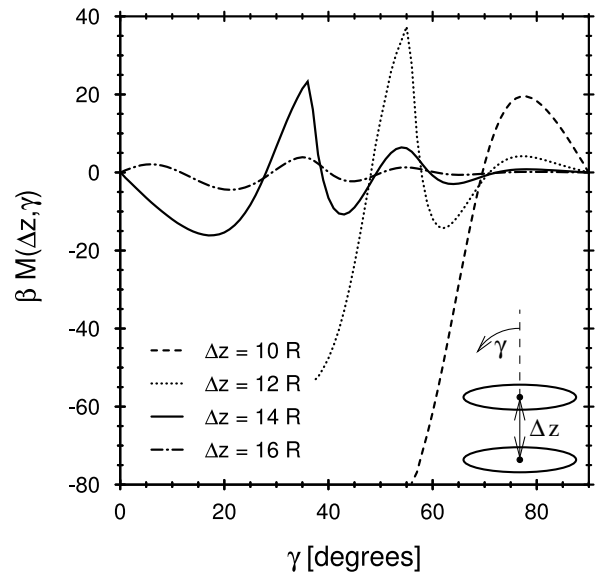}
\caption{\label{fig:oblatetorque}
Entropic torque between two oblate ellipsoids with half-axes $(10,10,4)R$
immersed in a solvent of hard spheres with radius $R$ as sketched in the
inset. The angle $\gamma$ parameterizes the rotation of the inserted ellipsoid
and $\Delta z$ denotes the distance between the centers. The ellipsoids are
aligned such that the vector connecting their centers runs through the north
poles of the fixed (lower) ellipsoid. For this setup the entropic torque $M$
is given by $M=- \partial W / \partial \gamma$ \cite{Roth02a}. If $\Delta z >
8R$, the inserted (upper) ellipsoid can be rotated as indicated in the
inset and experiences an entropic torque. Positive values of $M$ mean that the
ellipsoid is pushed towards larger values of $\gamma$. The
curves end if the two ellipsoids overlap due to the geometrical
constraint. Note that the visible discontinuities in the first derivative of
the curves for $\Delta z =12 R$ and $\Delta z = 14 R$ are not artifacts. They
occur if both ellipsoids come so close that just a single small sphere with
radius $R$ fits in between their surfaces. For $\Delta z=16 R$ the distance
between the ellipsoids is large enough so that this effect does not occur.}  
\end{figure}

As pointed out earlier, for non-spherical objects it is also possible to keep
the distance between their centers constant and change the relative
orientation. From the corresponding change in the depletion potential we can
obtain the entropic torque \cite{Roth02a} acting on the inserted particle. For
the same geometrical setup as in Fig.~\ref{fig:twooblates} we calculate the
torque $M$ for center to center separations $\Delta z/R=10$, 12, 14, and
16. Due to the symmetry of this setup the entropic torque acting on the
inserted ellipsoid, relative to its center, is parallel to the rotation
$\omegav$, i.e., ${\bf M}(\Delta z,\gamma)=M(\Delta z,\gamma) {\bf n}_\gamma$,
with ${\bf n}_\gamma=\omegav/\omega$ and
\begin{equation}
M(\Delta z,\gamma) = - \frac{\partial W(\Delta z,\gamma)}{\partial \gamma}.
\end{equation}
The symmetry of the problem leads to $M(\Delta z,\gamma)=0$ for $\gamma=0$ and
$90^\circ$. A positive value of the torque acts on the inserted ellipsoid as 
to increase the angle $\gamma$ (rotating it towards an orientation parallel to 
the fixed ellipsoid), while a negative value of $M$ leads to a decrease of 
$\gamma$ (rotating it towards an orientation normal to the fixed ellipsoid). 
Some typical examples of the torque as a function of
$\gamma$ for various values of $\Delta z$ are shown in 
Fig.~\ref{fig:oblatetorque}. For small values of $\Delta z$ the amplitude of 
the torque is largest; however, the hard-core interaction prohibits small 
values of $\gamma$ due to geometrical constraints. For larger separations 
$\Delta z$ between the ellipsoids, the accessible range of values of 
$\gamma$ increases until finally the inserted ellipsoid can rotate freely 
without encountering the fixed ellipsoid. Due to the symmetry of the problem 
the torque vanishes for  $\gamma=90^\circ$ as well as for $\gamma=0$, provided
this orientation is accessible.  

If for any relative orientation of the ellipsoids the minimal distance between 
the surfaces equals the diameter  $2 R$ of a solvent spheres, the torque
exhibits a cusp, as can be seen in the cases of $\Delta z > 10 R$. For 
$\Delta z = 10 R$ this orientation occurs at $\gamma=90^\circ$. Accordingly,
for sufficiently large $\Delta z$ no such cusps occur.

All these feature of the torque (see Fig.~\ref{fig:oblatetorque}) are similar
to those of the entropic torque acting on a spherocylinder close to a planar
wall \cite{Roth02a}.

From these results, one can speculate how two freely floating ellipsoids most
likely would approach each other. Similar to the case of a spherocylinder
close to a planar wall \cite{Roth02a} we find that the ellipsoids prefer to be
parallel once they touch each other. This configuration corresponds to
$\gamma=90^\circ$ in Fig.~\ref{fig:twooblates}. In order to be able to overcome
potential barriers while approaching, it is easier for the ellipsoids to
approach with a relative orientation different from the parallel
configuration. By first seeking contact in regions of high curvature, and then
by adapting the orientation such that regions of low curvature get in contact
with each other, potential barriers, which have to be overcome, are
significantly reduced and the minimum of the depletion potential can be
reached. This interpretation agrees with observations reported in 
Refs.~\cite{Kinoshita04,Kinoshita06,Li05} where the depletion potential
between two spherocylinder was studied.

\section{Summary and Conclusions}
\label{sec:conclusion}

We have extended our density functional theory approach for calculating
depletion potentials \cite{Roth00} to the effective interactions between
generally shaped convex particles with surface curvatures which vary smoothly. 
As in the previously studied \cite{Roth00} geometrically simple case of two
spherical particles, it is most efficient to carry out the calculation in
two steps. In the first step we describe the structure of the solvent close to
a fixed object. We have shown that a curvature expansion of the density
profile, or equivalently of the derivatives $\Psi_\alpha$ of the free energy 
density [Eq.~(\ref{cepsi})] provides an efficient tool. By
employing the curvature expansion we implicitly assume that the mean and
Gaussian curvatures of the surface of the solute vary smoothly across its
surface. Discontinuities in the curvatures, such as those observed at sharp
edges or close to the spherical caps of a spherocylinder cannot be captured
fully by the curvature expansion in its present form (see
Fig.~\ref{fig:compare}).

In a second step we insert the second particle into the inhomogeneous
solvent. To this end we employ an extension of fundamental measure theory to
non-spherical particles \cite{Rosenfeld94,Rosenfeld95}, which was already
used in the study of a spherocylinder close to a planar wall \cite{Roth02b}. 
Using this approach we have studied the depletion potential between
one ellipsoid and a big sphere immersed in a hard-sphere solvent. In this
case we could test the accuracy of our approach by carrying out the
calculations in two different ways, which employ the various approximations of
our approach independently. We have found excellent agreement between the two
routes (see Fig.~\ref{fig:test}) which provides confidence for the scheme
used. Furthermore, we have studied the depletion potential between two equally
sized oblate ellipsoids. For this  case we have illustrated the potential of
our approach. From the resulting depletion potential (see
Figs.~\ref{fig:twooblates} and \ref{fig:oblatesangle}) one can calculate the
entropic force acting on the centers of the solutes as well as the entropic
torque (Fig.~\ref{fig:oblatetorque}). This provides a picture for the likely
pathway of how two freely floating ellipsoids approach each other under the
action of entropic forces

Besides the application to colloidal mixtures of non-spherical objects and
spheres, this approach should prove useful to studying biological inspired
model key and lock systems \cite{Kinoshita02}, for which depletion
interactions between non-spherical objects and geometrically structured
substrates are considered. The non-spherical objects and the substrates
display a perfect geometrical match, similar to biological macromolecules
(key) which can form chemical bonds with a cavity (lock) only if they are
sufficiently close together. The issue is how the key can be guided into the
lock using a robust, chemically unspecific mechanism. Our present analysis
provides first steps towards addressing this issue.

\newpage

\end{document}